\documentclass[11pt]{article}
\usepackage{epsfig,amsmath,amscd,amssymb,graphicx}

\begin{document}
\thispagestyle{empty}

\newcommand {\beq}{\begin{eqnarray}}
\newcommand {\eeq}{\end{eqnarray}}
\newcommand {\non}{\nonumber\\}
\newcommand {\eq}[1]{\label {eq.#1}}
\newcommand {\defeq}{\stackrel{\rm def}{=}}
\newcommand {\gto}{\stackrel{g}{\to}}
\newcommand {\hto}{\stackrel{h}{\to}}
\newcommand {\1}[1]{\frac{1}{#1}}
\newcommand {\2}[1]{\frac{i}{#1}}
\newcommand {\thb}{\bar{\theta}}
\newcommand {\ps}{\psi}
\newcommand {\psb}{\bar{\psi}}
\newcommand {\ph}{\varphi}
\newcommand {\phs}[1]{\varphi^{*#1}}
\newcommand {\sig}{\sigma}
\newcommand {\sigb}{\bar{\sigma}}
\newcommand {\Ph}{\Phi}
\newcommand {\Phd}{\Phi^{\dagger}}
\newcommand {\Sig}{\Sigma}
\newcommand {\Phm}{{\mit\Phi}}
\newcommand {\eps}{\varepsilon}
\newcommand {\del}{\partial}
\newcommand {\dagg}{^{\dagger}}
\newcommand {\pri}{^{\prime}}
\newcommand {\prip}{^{\prime\prime}}
\newcommand {\pripp}{^{\prime\prime\prime}}
\newcommand {\prippp}{^{\prime\prime\prime\prime}}
\newcommand {\pripppp}{^{\prime\prime\prime\prime\prime}}
\newcommand {\delb}{\bar{\partial}}
\newcommand {\zb}{\bar{z}}
\newcommand {\mub}{\bar{\mu}}
\newcommand {\nub}{\bar{\nu}}
\newcommand {\lam}{\lambda}
\newcommand {\lamb}{\bar{\lambda}}
\newcommand {\kap}{\kappa}
\newcommand {\kapb}{\bar{\kappa}}
\newcommand {\xib}{\bar{\xi}}
\newcommand {\ep}{\epsilon}
\newcommand {\epb}{\bar{\epsilon}}
\newcommand {\Ga}{\Gamma}
\newcommand {\rhob}{\bar{\rho}}
\newcommand {\etab}{\bar{\eta}}
\newcommand {\chib}{\bar{\chi}}
\newcommand {\tht}{\tilde{\th}}
\newcommand {\zbasis}[1]{\del/\del z^{#1}}
\newcommand {\zbbasis}[1]{\del/\del \bar{z}^{#1}}
\newcommand {\vecv}{\vec{v}^{\, \prime}}
\newcommand {\vecvd}{\vec{v}^{\, \prime \dagger}}
\newcommand {\vecvs}{\vec{v}^{\, \prime *}}
\newcommand {\alpht}{\tilde{\alpha}}
\newcommand {\xipd}{\xi^{\prime\dagger}}
\newcommand {\pris}{^{\prime *}}
\newcommand {\prid}{^{\prime \dagger}}
\newcommand {\Jto}{\stackrel{J}{\to}}
\newcommand {\vprid}{v^{\prime 2}}
\newcommand {\vpriq}{v^{\prime 4}}
\newcommand {\vt}{\tilde{v}}
\newcommand {\vecvt}{\vec{\tilde{v}}}
\newcommand {\vecpht}{\vec{\tilde{\phi}}}
\newcommand {\pht}{\tilde{\phi}}
\newcommand {\goto}{\stackrel{g_0}{\to}}
\newcommand {\tr}{{\rm tr}\,}
\newcommand {\GC}{G^{\bf C}}
\newcommand {\HC}{H^{\bf C}}
\newcommand{\vs}[1]{\vspace{#1 mm}}
\newcommand{\hs}[1]{\hspace{#1 mm}}
\newcommand{\al}{\alpha}
\newcommand{\be}{\beta}
\newcommand{\Lam}{\Lambda}

\newcommand{\kahler}{K\"ahler }
\newcommand{\con}[1]{{\Gamma^{#1}}}

\begin{flushright}
TIT/HEP--491 \\
{\tt hep-th/0302028} \\
February, 2003 \\
\end{flushright}
\vspace{3mm}

\begin{center}
{\Large\bf 
BPS WALL IN ${\cal N}=2$ SUSY \\
NONLINEAR SIGMA MODEL 
\\
WITH EGUCHI-HANSON 
MANIFOLD
}

\vskip 1cm

{\large Masato Arai~$^{a}$, Masashi~Naganuma~$^{b}$, \\
Muneto~Nitta~$^{c}$,~and~~  Norisuke Sakai~$^{b}$}


$^{a}$ 
  Institute of Physics, AS CR, 
  182 21, Praha 8, Czech Republic \\
$^{b}$  Department of Physics, 
Tokyo Institute of Technology, \\
Tokyo 152-8551, JAPAN 
\\
$^{c}$  Department of Physics, Purdue University, \\
West Lafayette, IN 47907-1396, USA
\end{center}

\vskip 2 cm

\begin{abstract}
BPS wall solutions are obtained for ${\cal N}=2$ supersymmetric 
 nonlinear 
sigma model with Eguchi-Hanson target manifold 
in a manifestly supersymmetric manner. 
The model is constructed by a massive hyper-K\"ahler quotient 
method both in the ${\cal N}=1$ superfield 
and in the ${\cal N}=2$ superfield (harmonic superspace). 
We describe the model in simple terms and give 
relations between various parameterizations 
which are useful to describe the model and the solution. 
Some more details 
can be found in our previous paper \cite{ANNS} [hep-th/0211103]. 
This article is 
dedicated to Professor Hiroshi Ezawa on the occasion of his 
seventieth birthday. 
\end{abstract}

\newpage

\section{Introduction}     

Supersymmetry (SUSY) has been a most promising guiding principle 
to construct realistic unified models beyond the standard 
model.\cite{DGSW} 
In recent years there have been vigorous studies on 
models with extra dimensions,\cite{LED,RS} where 
our world is assumed 
to be realized on an extended topological defects 
such as domain walls or various branes. 
Supersymmetry can be combined with this {\it brane-world scenario} 
and helps the construction of the extended topological defects. 

Solitons saturating an energy bound, 
called the BPS bound,\cite{BPS,WittenOlive} 
have played a crucial role also in 
non-perturbative studies of supersymmetric 
(SUSY) field theories in four dimensions.\cite{SeibergWitten}  
BPS domain walls are topological solitons of co-dimension one, 
which depend on one spatial coordinate and connect two 
SUSY vacua.
Since they preserve half of the original SUSY, 
they are called ${1 \over 2}$ BPS states.  
Such BPS domain walls were well studied 
in various models with global 
${\cal N}=1$ SUSY in four dimensions.\cite{CGR}  
Non-BPS multi-wall solutions were also studied to understand 
the SUSY breaking mechanism on the brane due to the coexistence 
of the other brane.\cite{MSSS}${}^-$\cite{SS} 
More recently we have constructed an exact BPS wall solution 
as well as non-BPS multi-wall solutions in the supergravity 
theory in four dimensions.\cite{EMSS} 
The intersections or junctions of domain walls 
preserve ${1 \over 4}$ of the original SUSY and have 
been discussed in ${\cal N}=1$ 
models in four 
dimensions.\cite{AbrahamTownsend}${}^-$\cite{GrSh} 

In order to consider models with extra dimensions, 
we need to discuss 
supersymmetric theories in spacetime with dimensions 
higher than four. 
They should have at least eight supercharges. 
The simplest field theory with eight SUSY is based on 
hypermultiplets containing only 
scalar and spinor as physical fields. 
Recently we have formulated ${1 \over 2}$ 
BPS domain walls in an eight SUSY model in four 
dimensions.\cite{ANNS} 
Moreover we have also succeeded in constructing the ${1 \over 2}$ BPS 
wall consistently in five-dimensional supergravity.\cite{AFNS} 
Before discussing the SUSY five-dimensional theories, 
it is useful to consider models with eight SUSY 
in four dimensions without gravity.

The rest of our paper 
is organized as follows. 
Sec.~2 explains how to obtain nonlinear sigma models of 
hypermultiplets with eight SUSY. 
Secs.3, 4 and 5 are devoted to 
${\cal N}=1$ superfield formulation of the model. 
In Sec.~3, we present the model 
using the $U(1)$ gauge field. 
We give the bosonic part of the action and 
eliminate auxiliary fields in the Wess-Zumino gauge. 
The hyper-\kahler quotient method 
in terms of the so-called moment map 
becomes very clear. 
In Sec.~4, we eliminate auxiliary superfields 
in the superfield level,  
taking a gauge compatible with SUSY
rather than the Wess-Zumino gauge. 
This has the advantage because we obtain 
the Lagrangian in terms of independent superfields.
In Sec.~5, we use the $O(2)$ gauge field to 
formulate the model instead of the $U(1)$ gauge field.
Sec.~6 is devoted to a brief review of harmonic superspace formalism 
(HSF). 
In Sec.~7, we formulate the model in HSF 
and eliminate auxiliary fields in the Wess-Zumino gauge. 
In Sec.~8, the constraints are solved 
by independent fields. 
We close our paper by Sec.~9, 
in which the BPS equation and 
the domain wall solution are given.

\section{${\cal N}=2$ Model with Hypermultiplets in $4$ Dimensions}
\label{sc:N2hyper}

If we take two free chiral scalar supermultiplets $\phi$ and $\chi$ 
with a complex mass term $m 
$ between them, 
they together become a free massive hypermultiplet with ${\cal N}=2$ 
SUSY \cite{Fayet} 
\begin{eqnarray}
 {\cal L} 
&=&  
 \left[ \phi^* \phi + \chi^* \chi \right]_{\theta^2\bar\theta^2}
 + \left( m \left[ \chi \phi \right]_{\theta^2} 
 + {\rm c}.{\rm c}. \right) \nonumber \\
&=&
-\partial_\mu \phi^{*} \partial^\mu \phi  
-\partial_\mu \chi^{*} \partial^\mu \chi  
+ F_\phi^{*}F_\phi 
+ F_\chi^{*}F_\chi 
 \nonumber \\
&+&
 \left(m (F_\phi \chi+F_\chi \phi) + {\rm c}.{\rm c}.\right)
+ 
 {\rm fermionic} \;\; {\rm terms}
\nonumber \\
&=&
-\partial_\mu \phi^{*} \partial^\mu \phi  
-\partial_\mu \chi^{*} \partial^\mu \chi  
- |m|^2\left(\phi^{*} \phi + \chi^{*} \chi \right)
\nonumber \\
&
+
& 
 {\rm fermionic} \;\; {\rm terms}
\;, 
\label{eq:free-massive}
\end{eqnarray} 
where complex conjugate is denoted as c.c.~, and 
the scalar components are denoted by the same letter as the 
superfields\footnote{
We follow mostly the notation of Ref.~\cite{WB}, 
except that $\mu, \nu, \dots$ denote space time in four dimensions, 
} 
 $\phi, \chi$.  
Since four 
real scalar fields Re$\phi$, Im$\phi$, Re$\chi$, and Im$\chi$ 
are symmetric, we can form three 
complex fields 
using any one of the fields, say Re$\phi$ 
 with any one of the other three fields : 
 Re$\phi+i$Re$\chi$, and  Re$\phi+i$Im$\chi$ 
beside the ordinary ${\rm Re}\phi + i {\rm Im}\phi=\phi$. 
These three complex structures are completely symmetric 
and serve as a characterization of ${\cal N}=2$ SUSY 
for hypermultiplets. 
It has been shown that any nonlinear sigma 
model consisting of hypermultiplets should have 
a triplet of complex structures, 
and the target manifold should be 
hyper-K\"ahler\cite{AF1} (HK) 
in contrast to 
K\"ahler of the ${\cal N}=1$ SUSY 
nonlinear sigma model.\cite{Zu} 

Theories with eight SUSY are so restrictive that 
the nontrivial interactions require the nonlinearity of 
kinetic term (nonlinear sigma model) 
if there are only hypermultiplets. 
In order to obtain a wall solution, we need to have a nontrivial 
 potential. 
In the case of ${\cal N}=2$ SUSY nonlinear sigma model containing 
only hypermultiplets, one can introduce a nontrivial 
potential which is the square of the Killing vector 
 acting on the HK manifold multiplied 
by a mass parameter. 
Moreover the Killing vector has to be 
holomorphic with respect to 
the three complex structures (tri-holomorphic) \cite{AF2}. 
These models are called ``massive HK nonlinear sigma models''.  

Let us now explain a mechanism to obtain a nontrivial potential 
as a Sherk-Schwarz reduction\cite{SherkSchwarz} 
from six or five  dimensions.\cite{SierraTownsend} 
It is usually best to start from a model in spacetime with 
maximal dimensions which is allowed by the postulated number 
of SUSY charges. 
In the present case of eight SUSY, we should consider hypermultiplets 
in six dimensions. 
Let us first illustrate the dimensional reduction by 
a free massless hypermultiplet in six dimensions, 
since a mass term is  forbidden by SUSY. 
If two (one) spatial dimensions are compactified, a nonvanishing momentum 
in these compactified dimensions gives a 
complex (real) mass term resulting in Eq.(\ref{eq:free-massive}). 
This mass parameter gives rise to a central term 
$Z = -i(\partial_5+i \partial_6)$ 
in the ${\cal N}=2$ SUSY algebra in four dimensions. 
In the case of a nonlinear sigma model with the target space metric 
$g_{ij*}$ in six dimensions, the kinetic term (of bosonic part of 
the Lagrangian) reads 
\begin{equation}
{\cal L}=-g_{ij*}\partial_M \phi^i \partial_N \phi^{j*} \eta^{MN} ,
\label{eq:NLSM6D}
\end{equation}
where $\eta^{MN}={\rm diag.}(-1,1,1,1,1)$, $M, N = 0, \cdots, 5$. 
Then we can twist the boundary condition 
for the compactified 
directions, say $x^4$ and $x^5$ 
using a Killing vector $k^i(\phi, \phi^*)$ for the isometry of 
the target space metric $g_{ij*}$ 
\begin{equation}
 -i(\partial_5+i \partial_6)\phi^i=\mu k^i(\phi, \phi^*), 
 \quad \mu \in {\bf C} . 
\end{equation}
Then we obtain a nontrivial potential term $V(\phi)$ from the Lagrangian 
(\ref{eq:NLSM6D}) 
\begin{equation}
{\cal L}=-g_{ij*}\partial_\mu \phi^i \partial_\nu \phi^{j*} \eta^{\mu\nu}
-V(\phi, \phi^*) ,
\end{equation}
\begin{equation}
V(\phi, \phi^*)=|\mu|^2 g_{ij*}k^i(\phi, \phi^*) k^{j*}(\phi, \phi^*) .
 \label{sherk}
\end{equation}
This is the Sherk-Schwarz dimensional reduction. 
Since theories with eight SUSY have three complex structures, 
the Killing vector $k^i$ has to be holomorphic with respect to 
all three complex structures (tri-holomorphic). 

\begin{figure}[t]
\begin{center}
\leavevmode
  \epsfysize=4.0cm
  \epsfbox{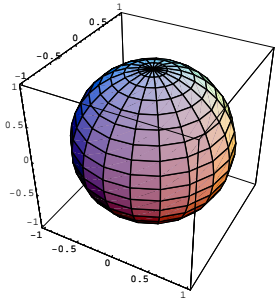} \\
\caption{
Sphere as a constrained subspace 
embedded in three-dimensional flat space. 
}
\label{fig:sphere}
\end{center}
\end{figure}

Many target space metrics can be 
embedded in higher dimensional flat space 
as illustrated by a 
sphere embedded in three 
dimensions in Fig.\ref{fig:sphere}. 
The nonlinear sigma models with these target space metrics 
can be realized by giving a constraint on hypermultiplets with 
minimal kinetic terms. 
One of the most convenient methods to impose the constraint 
is to introduce a vector multiplet without a kinetic term. 
If we integrate the vector multiplet, it acts as a Lagrange multiplier 
field to produce a constraint on hypermultiplets, 
resulting in a curved target manifold 
such as the Eguchi-Hanson manifold. 
In this process, the mass term automatically becomes a nontrivial 
potential which is a square of a Killing vector 
corresponding to the isometry of the resulting curved target space. 
Since the gauge field serves to identify the gauge orbit, 
the introduction of the gauge field without a kinetic term 
gives a quotient manifold. 
In particular, we call 
the method as hyper-K\"ahler quotient method, 
when the resulting manifold is hyper-K\"ahler. 
In our SUSY case, the curved manifold is a result of 
the constraint coming from integrating 
the auxiliary fields in the gauge supermultiplet 
as well as from the gauge orbit quotient.
Since we also have a mass term as a central extension of the SUSY algebra, 
this procedure is called the {\it massive hyper-K\"ahler 
quotient} method. 
In this way, we can understand the potential term as the 
square of the tri-holomorphic Killing vector and 
the mass parameter multiplying the potential 
as the central extension of the ${\cal N}=2$ SUSY algebra 
in four dimensions. 

There have been a number of works to study nonlinear sigma models 
 with eight supercharges.\cite{AF1,AF2,SierraTownsend}${}^{-}$\cite{HKLR} 
The massive nonlinear sigma
 model with 
 nontrivial K{\"a}hler metric as target space was
 studied, and BPS equations and 
 BPS solutions such as walls and junctions 
 were obtained.\cite{AT}${}^-$\cite{PT} 
Multi domain walls solution was also 
 obtained
 and the dynamics of those walls was examined.\cite{GTT,To}
Single or parallel domain walls in such models 
preserve ${1 \over 2}$ SUSY,\cite{AT,GTT} 
whereas their intersections preserve ${1 \over 4}$ SUSY.\cite{GTT1} 
In most papers, nonlinear sigma models were studied in
terms of component fields.
However, it is often useful to maintain as much SUSY as possible 
\cite{Sakamura}. 
Harmonic superspace 
 formalism (HSF) \cite{Ivanov} is most suited to maintain the SUSY 
 maximally, but there has been relatively few attempt to
 formulate the BPS equations and to obtain BPS 
 solutions in the HSF \cite{Zupnik} until our recent work.\cite{ANNS}

Our nonlinear sigma model can most easily be obtained 
by a quotient method in terms of a $U(1)$ 
vector multiplet without a kinetic term. 
In ${\cal N}=1$ formalism,
 the massless HK sigma model on $T^*{\bf C}P^{n}$ 
 was obtained as the HK quotient.\cite{RT,LR}
The massive HK quotient 
 was obtained in component level.\cite{To} 
The massless model with the Eguchi-Hanson target 
manifold ($T^*{\bf C}P^{1}$) \cite{EH} has been constructed in 
 ${\cal N}=2$ formalism,\cite{ivanov-EH} and 
 its central extension was analysed.\cite{Ketov}
In order to obtain also a potential term, 
we need to perform a quotient method for a massive 
hypermultiplet charged under the $U(1)$ vector multiplet.\cite{ANNS} 
When we are writing this up this work, 
another interesting work appeared discussing various wall and 
flux tube solutions in similar ${\cal N}=2$ models.\cite{ShifmanYung}

\section{Eguchi-Hanson  Nonlinear Sigma Model 
in ${\cal N}=1$ Superfields in the $U(1)$ Basis}
\label{sc:N1U1}

The massive hyper-K\"ahler quotient method 
for the massive Eguchi-Hanson \cite{EH} nonlinear sigma model 
requires two hypermultiplets : $\phi, \chi$ as doublets. 
The doublet $\phi (\chi)$ has charge $+1 (-1)$ 
under an ${\cal N}=2$ $U(1)$ vector 
multiplet $(V, \Sigma)$ without a kinetic term 
which serves as a Lagrange multiplier constraining 
hypermultiplets to form a four-dimensional (in terms of 
number of real degrees of freedom) target manifold. 
Here $V$ and $\Sigma$ are vector and chiral superfields
in ${\cal N}=1$ superfield formalism, respectively. 
Representing two doublets $\phi, \chi$ as column vectors, 
the action is given in terms of ${\cal N}=1$ superfields 
as \cite{ANNS} 
\begin{equation}
 {\cal L} 
=
 \left[ {\rm e}^V \phi^\dagger\phi 
+ {\rm e}^{-V} \chi^\dagger\chi - c V
\right]_{\theta^2\bar\theta^2} 
+\left(\left[\Sigma (\chi^T 
\phi - b)
+{\mu \over 2} \chi^T \sigma_3 \phi \right]_{\theta^2}
         + {\rm c.c.}\right) ,
\label{linear}
\end{equation}
where we have absorbed a common mass of hypermultiplets into the field 
$\Sigma$ and denote $\mu$ as a complex parameter for the mass splitting. 
The electric and magnetic Fayet-Iliopoulos (FI) parameters 
are denoted as $c\in {\bf R}, b \in {\bf C}$. 
We see below that these parameters become the value 
of the triplet of the moment map for the $U(1)$ gauge symmetry.

In the limit of $\mu=0$, the model has a global (flavor) symmetry 
$U(2) = SU(2)\times U(1)_{\rm A}$ defined by 
\begin{equation}
\phi \to \phi' = g \phi, \; \;
\chi \rightarrow \chi' = g^{*}\chi, \quad 
\Sigma \rightarrow \Sigma'= \Sigma, \qquad g \in U(2), 
\label{eq:globalSU2}
\end{equation}
and, in the case of $b=0$, it has the additional 
$U(1)_{\rm D}$ symmetry
\begin{equation}
 \phi \rightarrow \phi' = {\rm e}^{i\theta_{\rm D}}\phi, \qquad 
 \chi \rightarrow \chi' = {\rm e}^{i\theta_{\rm D}}\chi, \qquad 
 \Sigma \rightarrow \Sigma'= {\rm e}^{-2i\theta_{\rm D}}\Sigma\;.  
\label{eq:globalU1D}
\end{equation}
The $U(2)$ symmetry is consistent with the ${\cal N}=2$ SUSY, 
but the $U(1)_{\rm D}$ symmetry is only with 
the manifest ${\cal N}=1$ SUSY. 
The mass splitting parameter breaks this 
$U(2) [\times U(1)_{\rm D}]$ 
global symmetry (for $b=0$) down to 
$U(1)\times U(1) \in U(2)$ defined by 
$\phi \rightarrow {\rm e}^{i\theta_1 {\bf 1}+i\theta_2\sigma_3}\phi,  
\; \; \chi \rightarrow  {\rm e}^{-i\theta_1 {\bf 1}-i\theta_2\sigma_3}\chi$, 
with $\Sigma$ unchanged. 
The $U(1)$ subgroup parametrized by $\theta_1$ is gauged. 
Since this mass splitting parameter affects only the potential term 
without affecting the kinetic term, 
the curved target manifold has Killing vectors for the isometry 
 $SU(2) [\times U(1)_{\rm D}$ for $b=0$]. 
The $U(1)_{\rm A}$ symmetry of 
 $U(2)=SU(2)\times U(1)_{\rm A}$ is gauged away and 
is absent in the target manifold.  
However, only the Killing vectors for the isometry $SU(2)$ 
are consistent with the ${\cal N}=2$ SUSY and hence tri-holomorphic. 
The Killing vector for the $U(1)_{\rm D}$ isometry for $b=0$  
is holomorphic,
 but not tri-holomorphic.  
Since we have introduced the mass parameter through the $\sigma_3$ 
generator, we will eventually obtain the potential term 
which is a square of the tri-holomorphic Killing vector 
corresponding to $\sigma_3$, after eliminating the vector multiplet. 

In the Wess-Zumino gauge, the bosonic action is given by 
\begin{equation}
 {\cal L}_{\rm boson} = 
 {\cal L}_{\rm kin} + 
 {\cal L}_{\rm constr} + 
 {\cal L}_{\rm pot},  
\label{eq:N1U1boson}
\end{equation}
\begin{eqnarray}
 {\cal L}_{\rm kin} 
&
 = 
& 
-\left|\left(\partial_\mu +{i \over 2} v_\mu\right)\phi\right|^2
-\left|\left(\partial_\mu -{i \over 2} v_\mu\right)\chi\right|^2
\nonumber \\
&=& 
-|\partial_\mu \phi|^2
-|\partial_\mu \chi|^2
+ \frac{i}{2} v^\mu \left(
\phi^\dagger \overset{\leftrightarrow}{\partial_\mu}\phi 
 -\chi^\dagger \overset{\leftrightarrow}{\partial_\mu}\chi 
\right) 
\nonumber \\
&
 -
& 
\frac{1}{4}v^\mu v_\mu\left(\phi^\dagger \phi+ \chi^\dagger \chi \right) 
\end{eqnarray}
where $\phi^\dagger \overset{\leftrightarrow}{\partial_\mu}{\phi} 
\equiv \phi^\dagger (\partial_\mu \phi) 
- (\partial_\mu \phi^\dagger) \phi$. 
\begin{equation}
 {\cal L}_{\rm constr} = 
{D \over 2}\left(\phi^\dagger \phi -\chi^\dagger \chi-c \right) 
+
\left(F_\Sigma \left(\chi^T\phi - b\right) +
{\rm c.c.}\right), 
\label{eq:N1U1constr-lag}
\end{equation}
\begin{eqnarray}
 {\cal L}_{\rm pot} & = & 
F_\phi^\dagger F_\phi + F_\chi^\dagger F_\chi 
\nonumber \\
&
+
& 
\left(F_\chi^T \left(\Sigma {\bf 1} +{\mu \over 2}\sigma_3\right) \phi 
+\chi^T \left(\Sigma {\bf 1}+{\mu \over 2}\sigma_3\right) F_\phi 
 +
{\rm c.c.}\right)
\nonumber \\
&
\equiv
& 
 -V(\phi,\chi,\Sigma).
\label{eq:N1U1pot}
\end{eqnarray}

The equation of motion for gauge field $v_\mu$ in (\ref{eq:N1U1boson}) 
allows us to write gauge field $v_\mu$ in terms of scalar fields 
\begin{equation}
v_\mu = { i\left(
\phi^\dagger \overset{\leftrightarrow}{\partial_\mu}\phi 
 -\chi^\dagger \overset{\leftrightarrow}{\partial_\mu}\chi 
\right)  
\over \phi^\dagger \phi + \chi^\dagger \chi}.  
  \label{U(1)quotient}
\end{equation}
If we eliminate the gauge field by this algebraic equation of 
motion, we obtain the kinetic term for hypermultiplets as 
\begin{eqnarray}
 {\cal L}_{\rm kin}  =  
-|\partial_\mu \phi|^2
-|\partial_\mu \chi|^2
+ \frac{\left| i\left(
\phi^\dagger \overset{\leftrightarrow}{\partial_\mu}\phi 
 -\chi^\dagger \overset{\leftrightarrow}{\partial_\mu}\chi 
\right)  \right|^2}
{ 4(\phi^\dagger \phi + \chi^\dagger \chi)
}. 
\label{eq:N1U1kin}
\end{eqnarray}

If we integrate the Lagrange multiplier fields 
$D$ and $F_\Sigma$ in (\ref{eq:N1U1constr-lag}), 
we obtain two constraints 
\begin{equation}
\phi^\dagger \phi -\chi^\dagger \chi = c, 
\qquad 
\chi^T\phi = b  .
\label{eq:N1U1constr}
\end{equation}
The Lagrangian (\ref{eq:N1U1pot}) gives algebraic 
equations of motion for 
the auxiliary fields $F_\phi, F_\chi$ 
\begin{eqnarray}
F_\phi^\dagger = -\chi^T\left(\Sigma {\bf 1} +{\mu \over 2} \sigma_3\right), 
\qquad 
F_\chi^\dagger = - \phi^T\left(\Sigma {\bf 1} +{\mu \over 2} \sigma_3\right).  
\end{eqnarray}
After eliminating the auxiliary fields $F_\phi, F_\chi$ by these algebraic 
equations of motion,  
we obtain 
the potential term 
\begin{eqnarray}
V(\phi,\chi,\Sigma) &=& |F_\phi|^2+|F_\chi|^2 \nonumber \\
     &=& \left|\left(\Sigma {\bf 1} +{\mu \over 2} \sigma_3\right)\phi\right|^2 
     + \left| \left(\Sigma {\bf 1} +{\mu \over 2} \sigma_3\right)\chi\right|^2 
 \nonumber \\  
    &=& \left(\left|{\mu \over 2}\right|^2+|\Sigma|^2\right)
\left(\left|\phi\right|^2+\left|\chi\right|^2\right)
+\left({\mu^* \over 2}\Sigma + {\mu \over 2}\Sigma^*\right)
\left(\phi^\dagger \sigma_3 \phi + \chi^\dagger \sigma_3 \chi \right) 
 \nonumber \\  
&
=
&
\left|{\mu \over 2}\right|^2 
 \left[ |\phi|^2 + |\chi|^2 
- {\left(\phi^\dagger \sigma_3 \phi + \chi^\dagger \sigma_3 \chi \right)^2 
\over  |\phi|^2 + |\chi|^2 }
\right] .
\label{eq:N1U1pot2}
\end{eqnarray}
In the last line, we have eliminated the scalar field 
$\Sigma$ in the ${\cal N}=2$ 
vector multiplet $(V, \Sigma)$ by its algebraic equation 
of motion 
\begin{equation}
\Sigma = -{\mu \over 2}{\phi^\dagger \sigma_3 \phi 
+ \chi^\dagger \sigma_3 \chi \over 
|\phi|^2+|\chi|^2 } .
\end{equation}

In Eq.~(\ref{eq:N1U1constr}) 
the left hand sides constitute 
the triplet of the moment map (Killing potential) 
for the $U(1)$ gauge symmetry. 
Hence we see that these values are fixed to
the FI parameters by integrating the auxiliary fields $D$ and $F_{\Sigma}$, 
and that the hyper-\kahler quotient is obtained 
together with the $U(1)$ quotient (\ref{U(1)quotient}). 
In the limit of $b=c=0$ the singularity appears 
and the manifold becomes 
the orbifold ${\bf C}^2/{\bf Z}_2$, 
whereas the non-zero values of $b$ and $c$ 
resolve the orbifold singularity 
through the deformation of the complex structure 
and the blow up, respectively.

\section{
Eguchi-Hanson Nonlinear Sigma Model 
After Integrating Vector Multiplet 
in the $U(1)$ Basis }
\label{sc:N1U1NLSM}

Instead of taking the Wess-Zumino gauge in the component level, 
we can eliminate the auxiliary superfields 
$V$ and $\Sigma$ directly 
in the superfield formalism. 
Their equations of motion read from Eq.(\ref{linear}) as 
\begin{eqnarray}
 && {\partial {\cal L} \over \partial V} 
 = {\rm e}^V |\phi|^2 - {\rm e}^{-V} |\chi|^2 - c = 0 \; , 
   \label{EOM-V}\\
 && {\partial {\cal L} \over \partial \Sigma} 
 = \chi^T  \phi - b = 0\;,
   \label{EOM-sig}
\end{eqnarray}
in which $V$ can be solved immediately to give 
${\rm e}^V = (c \pm \sqrt {c^2 + 4|\phi|^2 |\chi|^2})/ 2|\phi|^2$. 
We thus obtain the K\"ahler potential 
\begin{equation}
 K = c \sqrt{1 + {4 \over c^2} |\phi|^2 |\chi|^2 } 
   - c \log \left(1 + \sqrt{1 
       + {4 \over c^2} |\phi|^2 |\chi|^2 } \right)
   + c \log |\phi|^2 \;, \label{kahler}
\end{equation}
where we have chosen the plus sign of the solution 
for the positivity of the metric. 

Fixing the complexified $U(1)$ gauge symmetry and solving 
(\ref{EOM-sig}), 
we can obtain the Lagrangian of the nonlinear sigma model
in terms of independent superfields. 
We have presented some gauge fixing \cite{ANNS} 
applied to $T^*{\bf C}P^n$ for general $n$. 
Here we give a more symmetric gauge for $T^* {\bf C}P^1$
in the case of $b=0$. 
In this case, we can fix the gauge as ${\chi_1 / \phi_2} = 1$,  
and (\ref{EOM-sig}) can be solved as 
$\phi^T = (x,y)$ and $\chi^T = (y , -x)$ 
(and hence $|\phi|^2 = |\chi|^2 = |x|^2 +|y|^2$).  
The K\"ahler potential becomes 
\cite{GP}  
\begin{equation}
 K = c\sqrt{1 + {4\over c^2} |\phi|^4} 
   - c\log\left( {1 + \sqrt{1 + {4\over c^2} |\phi|^4} 
                 \over |\phi|^2 } \right) \;,
\end{equation} 
and the superpotential is given by 
\begin{equation}
W = \mu x y \;. 
\end{equation} 
The metric and its inverse can be calculated to give 
\begin{eqnarray}
 && g_{ij^*} = 
 { c\over |\phi|^4 \sqrt{1+{4\over c^2} |\phi|^4} } 
\left(\begin{array}{cc}
 |y|^2 + {4\over c^2}|\phi|^6 & - x^* y \\ 
            - y^* x & |x|^2 + {4\over c^2}|\phi|^6 
	      \end{array}\right)
         \; , \nonumber \\
 && g^{ij^*} = 
 { c\over 4 |\phi|^4 \sqrt{1+{4\over c^2} |\phi|^4} } 
\left(\begin{array}{cc}
 |x|^2 + {4\over c^2}|\phi|^6 &   y^* x \\ 
            x^* y & |y|^2 + {4\over c^2}|\phi|^6 
	      \end{array}\right)
          \;, 
\end{eqnarray}
where $g_{ij^*}=\partial^2 K/\partial\phi^i\partial\phi^{j*}$ and
 $\phi^i=(x,y)$.
The scalar potential can be calculated as 
\beq
  V = g^{ij^*} \del_i W \del_{j^*} W^* 
     = { c |\mu|^2\over |\phi|^4 \sqrt{1+{4\over c^2} |\phi|^4} } 
        \left({1 \over c^2} |\phi|^8 + |x|^2 |y|^2 \right)  \;.
\eeq

The manifold admits 
the tri-holomorphic isometry $SU(2)$, 
defined in Eq.(\ref{eq:globalSU2}).\footnote{
The $SU(2)$ transformation law of 
$\phi^i$ as the coordinates of 
the quotient target manifold 
is unchanged from the one in Eq.~(\ref{eq:globalSU2}) 
and hence is still {\it linear}, 
because the gauge fixing condition 
is invariant under the $SU(2)$ action. 
This is the advantage of our gauge fixing condition. 
To define the $SU(2)$ action 
in the cases of the other gauge condition~\cite{ANNS}, 
we need an appropriate $U(1)$ gauge action 
to compensate the variation of the gauge condition, 
which makes the transformation law of $\phi^i$ {\it nonlinear}.

The diagonal $U(1)_{\rm D}$ isometry defined in Eq.(\ref{eq:globalU1D}) 
of $U(2)\times U(1)_{\rm D}=SU(2)\times U(1)_{\rm A}\times U(1)_{\rm D}$ 
isometry is holomorphic but not tri-holomorphic.
}
The Killing vectors for this action  
$k^i_A = \1{\epsilon} \delta_{\epsilon} \phi^i 
= {i \over 2} (\sigma_A)^i{}_j \phi^j $ ($A=1,2,3$) 
are given as 
\begin{equation}
 (k_1,k_2,k_3) 
 = {i \over 2} \left(  
  \left(\begin{array}{c}
          y \\
          x
	\end{array}\right), 
  \left(\begin{array}{c}
          -iy \\
           ix
	\end{array}\right), 
  \left(\begin{array}{c}
           x \\
          -y
	\end{array}\right)
 \right) \;. 
\end{equation}
The Killing potential $D_A(\phi,\phi^*)$ for these vectors, 
defined by $k_A^i = i g^{ij^*} \del_{j^*} D_A$,   
are given as~\cite{HKLR}
\begin{equation}
 (D_1,D_2,D_3) 
\! = \!
{c \over 2} {\sqrt{1+ {4\over c^2}|\phi|^4} \over |\phi|^2} 
   \left( xy^* + yx^*, i(xy^* - y x^*), |x|^2 - |y|^2 \right) .
\end{equation}
Using these geometric quantities, 
the scalar potential can be rewritten 
 by the square of the Killing vector 
\begin{equation}
  V = |\mu|^2 g_{ij^*} k_3^i k_3^{*j}  
    = |\mu|^2 g^{ij^*} \del_i D_3 \del_{j^*} D_3 
\end{equation}
as was shown in Eq.~(\ref{sherk}).
It is now manifest that only the action of 
$k_3$ among three Killing vectors 
preserves the potential term and 
therefore is the symmetry of the whole Lagrangian 
as expected from the mass term in (\ref{linear}).  
The vacua are fixed points of 
the Killing vectors $k_3$ 
or the critical points of 
the Killing potential $D_3$.

\section{Eguchi-Hanson  Nonlinear Sigma Model  
 in ${\cal N}=1$ Superfields in the $O(2)$ Basis}
\label{sc:N1O2}

It is also useful to rewrite the model in terms of $O(2)$ gauge group 
instead of $U(1)$, since $O(2)$ basis 
is most frequently employed in the harmonic superspace 
formalism as given in the next section. 
Introducing  $O(2)$ doublets $\tilde \phi^a$ as a column vector and 
$\tilde \chi_a$ as a row vector, 
superspace action in the $O(2)$ basis is given by 
\begin{eqnarray}
 {\cal L}^{O(2)}
&=& \left[\tilde \phi_a^\dagger ({\rm e}^{VT})^a{}_b \tilde \phi^b
 + \tilde \chi_a ({\rm e}^{-VT})^a{}_b \tilde \chi^{b\dagger}-cV 
\right]_{\theta^2 \bar\theta^2}
  \nonumber \\
         & &~~~~~~~~+\left(\left[
     \Sigma \left(\tilde \chi_a T^a{}_b \tilde \phi^b - b\right) 
+{\mu \over 2}\tilde \chi_a \tilde \phi^a\right]_{\theta^2}
                 +{\rm c.c.} \right),
\label{eq:N1O2lagrangian}
\end{eqnarray}
where the hermitian $O(2)$ generator is given from a diagonal 
generator $\sigma_3$ by 
\begin{equation}
T \equiv \sigma_2 
=
{\rm e}^{i{\pi \sigma_1 \over 4}} \sigma_3
{\rm e}^{-i{\pi \sigma_1 \over 4}} 
=
{1 + i \sigma_1 \over \sqrt2} \sigma_3
{1 - i \sigma_1 \over \sqrt2}
.
\end{equation}
In order to establish a relation between superspace action in the 
$U(1)$ and $O(2)$ bases, it is convenient to define fields 
$\tilde \phi', \tilde \chi'$ with 
definite $U(1)$ charge by means of a rotation from the $O(2)$ 
basis 
\begin{equation}
\tilde \phi' \equiv 
{1 - i \sigma_1 \over \sqrt2} \tilde \phi, 
\qquad 
\tilde \chi' \equiv  \tilde \chi 
{1 - i \sigma_1 \over \sqrt2} 
.
\end{equation}
In terms of these superfields, the action becomes 
\begin{eqnarray}
 {\cal L}^{O(2)}
&=& \left[\tilde \phi_a^{'\dagger} 
({\rm e}^{V \sigma_3})^a{}_b \tilde \phi^{'b}
 + \tilde \chi_a' 
({\rm e}^{V \sigma_3})^{a}{}_b \tilde \chi^{'b\dagger}-cV 
\right]_{\theta^2 \bar\theta^2}
  \nonumber \\
         & + & \left(\left[
     \Sigma \left(\tilde \chi_a' \left(\sigma_2\right)^a{}_b 
\tilde \phi^{'b} - b\right) 
+{\mu \over 2}\tilde \chi'_a \left(i\sigma_1\right)^a{}_b 
 \tilde \phi^{'a}\right]_{\theta^2}
                 +{\rm c.c.} \right).
\end{eqnarray}
By identifying the $O(2)$ and $U(1)$ gauge fields, 
we find that the superfields $\phi_{i}, \chi_{i}$ in the $U(1)$ 
basis should be related to the superfields 
$\tilde \phi^{'i}, \tilde \chi^{'}_i$ in the $O(2)$ 
basis as 
\begin{eqnarray}
 |\tilde \phi^{'1}|^2 +|\tilde \chi'_1|^2 
&
= 
&
 | \phi_{1}|^2 +|\phi_2|^2,  
\label{eq:correspondence1} \\
 |\tilde \phi^{'2}|^2 +|\tilde \chi'_2|^2 
&
= 
&
 |\chi_{1}|^2 +|\chi_2|^2 ,  
\label{eq:correspondence2} \\
 i\tilde \chi^{'}_2 \tilde \phi^{'1} 
-i\tilde \chi'_1 \tilde \phi^{'2} 
&=& 
  \chi_1 \phi_1 
+ \chi_2 \phi_2 , 
\label{eq:correspondence3}   \\
 i\tilde \chi^{'}_2 \tilde \phi^{'1} 
+i\tilde \chi'_1 \tilde \phi^{'2} 
&=& 
  \chi_1 \phi_1 
- \chi_2 \phi_2 
.
\label{eq:correspondence4}
\end{eqnarray}
The most general solution of the conditions 
(\ref{eq:correspondence1}) and 
(\ref{eq:correspondence2}) 
is given in terms of two unitary matrices $U, V$ 
\begin{eqnarray}
\left(\begin{array}{c}
	  \tilde \phi^{'1} \\
	  \tilde \chi'_1 
	      \end{array}\right)
&
=
& U 
\left(\begin{array}{c}
	   \phi_1 \\
       \phi_2 
	      \end{array}\right), 
\qquad 
U^\dagger U=1,
\\
\left(\begin{array}{c}
	  \tilde \phi^{'2} \\
	  \tilde \chi'_2 
	      \end{array}\right)
&
=
& V 
\left(\begin{array}{c}
	   \chi_1 \\
       \chi_2 
	      \end{array}\right), 
\qquad 
V^\dagger V=1.
\end{eqnarray}
The constraints (\ref{eq:correspondence3}) and 
(\ref{eq:correspondence4}) give conditions 
\begin{equation}
V^T \sigma_2 U=1, 
\qquad 
V^T i \sigma_1 U=\sigma_3, 
\end{equation}
respectively. 
The first condition gives $V^T= U^\dagger \sigma_2$. 
By substituting it to the second condition we obtain 
\begin{equation}
U^\dagger \sigma_3 U = \sigma_3. 
\end{equation}
The most general 
solution of these conditions is now given 
in terms of two arbitrary angle parameters $\alpha, \beta$ by 
\begin{equation}
U=
\left(\begin{array}{cc}
 {\rm e}^{i\alpha} & 0 \\
 0                 & {\rm e}^{i\beta} 
 \end{array}\right), 
\qquad 
V
=
- \sigma_2 
\left(\begin{array}{cc}
 {\rm e}^{-i\alpha} & 0 \\
 0                 & {\rm e}^{-i\beta} 
 \end{array}\right). 
\end{equation}
These angles $\alpha, \beta$ clearly represent the 
$U(1)\times U(1)$ symmetry of our Lagrangian. 
Therefore a general solution for the identification 
 of superfields $\phi_i, \chi_i$ in the $U(1)$ basis and 
 superfields $\tilde \phi^a, \tilde \chi_a$ in the $O(2)$ 
superfields is given by 
\begin{eqnarray}
\phi
&
=
&
\left(\begin{array}{c}
	   \phi_1 \\
       \phi_2 
	      \end{array}\right)
=
\left(\begin{array}{c}
{\rm e}^{-i\alpha}
	  \tilde \phi^{'1} \\
{\rm e}^{-i\beta}
	  \tilde \chi'_1 
	      \end{array}\right)
={1 \over \sqrt2}
\left(\begin{array}{c}
{\rm e}^{-i\alpha}
( \tilde \phi^{1}-i \tilde \phi^{2}) \\
{\rm e}^{-i\beta}
( \tilde \chi_1 -i \tilde \chi_2 ) 
	      \end{array}\right),
\\
\chi
&
=
&
\left(\begin{array}{c}
	   \chi_1 \\
       \chi_2 
	      \end{array}\right)
=
\left(\begin{array}{c}
i{\rm e}^{i\alpha}
	  \tilde \chi^{'}_2 \\
-i{\rm e}^{i\beta}
	  \tilde \phi^{'2} 
	      \end{array}\right)
={1 \over \sqrt2}
\left(\begin{array}{c}
{\rm e}^{i\alpha}
( \tilde \chi_{1}+i \tilde \chi_{2}) \\
{\rm e}^{i\beta}
(- \tilde \phi^1 -i \tilde \phi^2 ) 
	      \end{array}\right).
\end{eqnarray}
 From now on, we shall take $\alpha=\beta=0$ case 
as a representative choice 
which is given by  
\begin{equation}
\left(\begin{array}{c}
	   \phi_1 \\
       \phi_2 
	      \end{array}\right)
={1 \over \sqrt2}
\left(\begin{array}{c}
 \tilde \phi^{1}-i \tilde \phi^{2} \\
 \tilde \chi_1 -i \tilde \chi_2  
	      \end{array}\right), 
\qquad
\left(\begin{array}{c}
	   \chi_1 \\
       \chi_2 
	      \end{array}\right)
={1 \over \sqrt2}
\left(\begin{array}{c}
 \tilde \chi_{1}+i \tilde \chi_{2} \\
 -\tilde \phi^1 -i \tilde \phi^2  
	      \end{array}\right).
\end{equation}

Bosonic part of the Lagrangian in the $O(2)$ basis 
(\ref{eq:N1O2lagrangian}) is given in the Wess-Zumino gauge 
\begin{equation}
 {\cal L}_{\rm boson}^{O(2)} = 
 {\cal L}_{\rm kin}^{O(2)} + 
 {\cal L}_{\rm constr}^{O(2)} + 
 {\cal L}_{\rm pot}^{O(2)},  
\label{eq:N1O2boson}
\end{equation}
\begin{eqnarray}
{\cal L}_{\rm kin}^{O(2)}&=&
-\left|\left(\partial_\mu +{i \over 2} v_\mu T\right)
\tilde \phi\right|^2
-\left|\left(\partial_\mu +{i \over 2} v_\mu T\right)
\tilde \chi^T \right|^2
\nonumber \\
&
=
&
-|\partial_\mu \tilde \phi|^2
-|\partial_\mu \tilde \chi|^2
 +\frac{i}{2}v^\mu \left(
\tilde \phi^\dagger T 
 \overset{\leftrightarrow}\partial_\mu\tilde \phi 
+\tilde \chi T 
 \overset{\leftrightarrow}\partial_\mu\tilde \chi^\dagger 
\right)
\nonumber \\
&
 -
&
\frac{1}{4}v^\mu v_\mu\left(
\tilde \phi^\dagger 
\tilde \phi + 
\tilde \chi 
\tilde \chi^\dagger \right),
\end{eqnarray}
\begin{eqnarray}
 {\cal L}_{\rm constr}^{O(2)} 
&
=
& 
{D \over 2}\left(\tilde \phi^\dagger T \tilde \phi 
-\tilde \chi T \tilde \chi^\dagger -c \right) 
+
\left(
F_\Sigma \left(\tilde \chi T \tilde \phi - b\right) +
{\rm c.c.}\right),
\label{eq:N1O2constrlagr}
\end{eqnarray}
\begin{eqnarray}
 {\cal L}_{\rm pot}^{O(2)}  
&
=
&
 F_{\tilde \phi}^\dagger F_{\tilde \phi} 
+ F_{\tilde \chi} F_{\tilde \chi}^\dagger 
\nonumber \\
&
+
& 
\left(F_{\tilde \chi} \left(\Sigma T +{\mu \over 2}{\bf 1}\right) 
\tilde \phi 
+\tilde \chi \left(\Sigma T+{\mu \over 2}{\bf 1}\right) 
F_{\tilde \phi} 
 +
{\rm c.c.}\right)
\nonumber \\
&
\equiv 
&
 -V(\phi,\chi,\Sigma). \label{eq:N1O2pot}
\end{eqnarray}

The equation of motion for gauge field $v_\mu$ is given by 
\begin{equation}
v_\mu = { i\left(
\tilde \phi^\dagger T 
 \overset{\leftrightarrow}\partial_\mu\tilde \phi 
+\tilde \chi T 
 \overset{\leftrightarrow}\partial_\mu\tilde \chi^\dagger 
\right)  
\over 
\tilde \phi^\dagger \tilde \phi + 
\tilde \chi \tilde \chi^\dagger 
}.
\end{equation}
Eliminating the gauge field by this algebraic equation of 
motion, we obtain the kinetic term for hypermultiplets as 
\begin{eqnarray}
 {\cal L}_{\rm kin}  =  
-|\partial_\mu \tilde \phi|^2
-|\partial_\mu \tilde \chi|^2
+ \frac{\left| 
 i\left(
\tilde \phi^\dagger T 
 \overset{\leftrightarrow}\partial_\mu\tilde \phi 
+\tilde \chi T 
 \overset{\leftrightarrow}\partial_\mu\tilde \chi^\dagger 
\right) 
  \right|^2}
{ 4(\tilde \phi^\dagger \phi + \tilde \chi^\dagger \chi)
}. 
\label{eq:N1U1kin2}
\end{eqnarray}
Integrating the Lagrange multiplier fields 
$D$ and $F_\Sigma$ in (\ref{eq:N1O2constrlagr}), 
we obtain two constraints 
\begin{equation}
\tilde \phi^\dagger T \tilde \phi 
-\tilde \chi T \tilde \chi^\dagger -c
=0, 
\qquad 
\tilde \chi T \tilde \phi - b=0 .
\label{eq:N1O2constr}
\end{equation}
Eliminating 
the algebraic 
equations of motion for 
the auxiliary fields $F_\phi, F_\chi$ 
\begin{eqnarray}
F_\phi^\dagger = -
\tilde\chi \left(\Sigma T+{\mu \over 2}{\bf 1} \right)
, 
\qquad 
F_\chi^\dagger = - 
\left(\Sigma T +{\mu \over 2}{\bf 1} \right)\tilde \phi
.  
\end{eqnarray}
After eliminating the auxiliary fields $F_\phi, F_\chi$ by these algebraic 
equations of motion,  
we obtain 
the potential term 
\begin{eqnarray}
V(\tilde \phi, \tilde \chi, \Sigma) 
&=& |F_{\tilde \phi}|^2+|F_{\tilde \chi}|^2 \nonumber \\
 &=& \left|\left(\Sigma T +{\mu \over 2}{\bf 1} \right)\tilde \phi\right|^2 
 + \left|\tilde\chi \left(\Sigma T+{\mu \over 2}{\bf 1} \right)\right|^2 
 \nonumber \\  
    &=& \left(\left|{\mu \over 2}\right|^2+|\Sigma|^2\right)
\left(|\tilde \phi|^2+\left|\tilde \chi\right|^2\right)
+\left({\mu^* \over 2}\Sigma + {\mu \over 2}\Sigma^*\right)
\left(\tilde \phi^\dagger T \tilde \phi + 
\tilde \chi T \tilde \chi^\dagger \right) 
 \nonumber \\  
&
=
&
\left|{\mu \over 2}\right|^2 
 \left[ |\tilde \phi|^2 + |\tilde \chi|^2 
- {\left(\tilde \phi^\dagger T \tilde \phi 
+ \tilde \chi T \tilde \chi^\dagger \right)^2 
\over  |\tilde \phi|^2 + |\tilde \chi|^2 }
\right] .
\label{eq:N1O2pot2}
\end{eqnarray}
In the last line, we have eliminated the scalar field 
$\Sigma$ in the ${\cal N}=2$ 
vector multiplet $(V, \Sigma)$ by its algebraic equation 
of motion 
\begin{equation}
\Sigma = -{\mu \over 2}
{\tilde \phi^\dagger T \tilde \phi 
+ \tilde \chi T \tilde \chi^\dagger \over 
|\tilde \phi|^2+|\tilde \chi|^2 } .
\end{equation}

\section{
A Brief Survey of Harmonic Superspace Formalism 
}\label{sc:HSF}

Harmonic superspace is defined as 
 $(x^\mu,~\theta_i,~{\bar{\theta}}^i,u_i^{\pm})$ which is called
 the central basis.
The $u_i^{\pm}$ are called the harmonic variables which parameterize 
 the coset $SU(2)_R/U(1)_r\sim S^2$,
 where $i=1,2$ is $SU(2)_R$ index and $\pm$ denotes $U(1)_r$ charge.
The superfield in the Harmonic Superspace Formalism 
 (HSF) is not defined in the central basis 
 but in the subspace which is called the analytic subspace 
\begin{equation}
\{\zeta_A,u_i^{\pm}|x_A^\mu=x^\mu 
  - 2 i \theta^{(i}\sigma^\mu {\bar{\theta}}^{j)}u_{(i}^+u_{j)}^-,
  ~\theta^+ = \theta^i u_i^+,
  ~{\bar{\theta}}^+ = {\bar{\theta}}^i u_i^{+},
  ~u_i^{\pm}\},
\end{equation}
  where parentheses for indices $i,j$ mean symmetrization, for instance, 
\begin{equation}
 u_{(i}^+u_{j)}^-=(u_i^+ u_j^- + u_j^+ u_i^-)/2 .
\end{equation}
Hypermultiplet and vector multiplet superfields are defined as the
 function in the analytic subspace as $q^+(\zeta_A,u)$ and
 $V^{++}(\zeta_A,u)$, respectively, which are called 
the analytic superfields.  

To describe the real action in terms of the analytic superfield,
 the star conjugation must be introduced in addition to the
 usual complex conjugation.
The complex conjugation rules for the coefficients in the harmonic
 expansions $f^{i_1\cdots i_n}$ (see (\ref{eq:harmonicexp})), 
 the Grassmann variables 
 $\theta_{i\alpha}$ and the harmonic variables $u_i^{\pm}$
 are defined as
\begin{eqnarray}
 \overline{f^{i_1\cdots i_n}} 
  &\equiv & {\bar{f}}_{i_1\cdots i_n},
\qquad 
\overline{f_{i_1\cdots i_n}} 
   =  (-1)^n {\bar{f}}^{i_1\cdots i_n}, 
   \label{eq:realityHSF} \\
 \overline{\theta_{i\alpha}} & = &  {\bar{\theta}}_{\dot{\alpha}}^{i}, 
\qquad 
\overline{\theta_{\alpha}^i} = -{\bar{\theta}}_{\dot{\alpha}i},\\
\qquad 
 \overline{u^{+i}} & = & u_i^-,
\qquad 
\overline{u_i^+}=-u^{-i},
\end{eqnarray}
respectively.
The star conjugation rules are defined as
\begin{eqnarray}
 &(f^{i_1\cdots i_n})^* = f^{i_1\cdots i_n},& \\
 &(\theta_{\alpha}^i)^* = \theta_{\alpha}^i,& \\
 &(u^{+i})^{*} = u^{-i},~~(u_i^{+})^{*} = u_i^{-},~~
 (u^{-i})^{*} = -u^{+i},~~(u_i^{-})^{*} = - u_i^{+},&\\
 &(u_i^{\pm})^{**} = -u_i^{\pm}.&
\end{eqnarray}
Note that the star conjugate acts only on the quantity having $U(1)_r$ charge.
We write the combination of the complex and the star conjugation
 as
\begin{eqnarray}
 (\overline{q^+(\zeta_A,u)})^* \equiv \widetilde{q^+}(\zeta_A,u).
\end{eqnarray} 
The combined conjugation rules are defined by
\begin{eqnarray}
&\widetilde{f^{i_1\cdots i_n}}
  =\overline{f^{i_1\cdots i_n}} 
  \equiv  {\bar{f}}_{i_1\cdots i_n}, & \\
& {\widetilde{\theta^+}} = {\bar{\theta}}^+,~~ 
  ~~{\widetilde{\theta^-}} = {\bar{\theta}}^-,
  ~~\widetilde{{\bar{\theta}}^+} = -\theta^+,
  ~~\widetilde{{\bar{\theta}}^-} = -\theta^-, & \\
& (\widetilde{u_i^{\pm}}) = u^{\pm i},
  ~~(\widetilde{u^{\pm i}})=-u_i^{\pm}. &
\end{eqnarray}

The simple example of the real action is the free massless action 
 of the Fayet-Sohnius hypermultiplet;
\begin{eqnarray}
 S = -\int d\zeta_A^{(-4)} du~{\widetilde{\phi^+}}D^{++}\phi^+
\end{eqnarray}
 where $D^{++}$ is defined by 
\begin{eqnarray}
 D^{++}=\partial^{++}-2 i \theta^+ \sigma^\mu {\bar{\theta}}^+ 
        \partial_\mu^A - (\theta^{+2}\bar{Z}-{\bar{\theta}}^{+2}Z), 
        \label{covdhss}
\end{eqnarray}
with $Z=0$ for a massless hypermultiplet.
Suffix $A$ in the spacetime derivative $\partial_\mu^A$ denotes 
the variable appropriate in the analytic superspace,\cite{Ivanov2} 
$\partial^{++}$ is the harmonic differential defined by 
 $\partial^{++}=u_i^+\frac{\partial}{\partial u_i^-}$. 
 For details of notation in HSF, we refer to our paper \cite{ANNS} 
or a textbook.\cite{Ivanov2} 

The action is real in the sense of ordinary complex conjugation
 $\bar{S}=S$.
This property follows from the fact that 
 ${\widetilde{\widetilde{q^+}}}=-q^+$.

%
%
Analytic superfields for the hypermultiplet 
$q_a^+(x_A, \theta^{\pm}, u)$ can be 
expanded in powers of Grassmann numbers $\theta$ as 
\begin{eqnarray}
&& q_a^+(x_A, \theta^{\pm}, u) 
 = F_a^+(x_A, u) + \sqrt{2} \theta^+ \psi_a(x_A, u) 
 + \sqrt{2} {\bar{\theta}}^+ {\bar{\varphi}}_a(x_A, u) 
                        \nonumber \\
 &  +& i\theta^+ \sigma^\mu {\bar{\theta}}^+ A_{a\mu}^-(x_A, u)
 + \theta^+ \theta^+ M_a^-(x_A, u) 
 + {\bar{\theta}}^+{\bar{\theta}}^+ N_a^-(x_A, u) 
         \nonumber \\
&  +& \sqrt{2} \theta^+\theta^+{\bar{\theta}}^+ {\bar{\chi}}_a^{--}(x_A, u)
 + \sqrt{2} {\bar{\theta}}^+ {\bar{\theta}}^+\theta^+ \xi_a^{--}(x_A, u)
                        \nonumber \\
&  +& \theta^+ \theta^+ {\bar{\theta}}^+ {\bar{\theta}}^+ D_a^{---}(x_A, u),
                      \label{eq:exp1} 
\end{eqnarray}
where $a$ is a flavor index. 
Note that each component 
 in the hypermultiplet analytic superfield (\ref{eq:exp1}) 
 is a function of $x_A$, and the harmonic variables $u_i^{\pm}$. 
 Therefore it 
 includes infinite series of functions of $x_A$ 
 when expanded by the
 harmonic variables  $u_i^{\pm}$ (harmonic expansions), for instance,
\begin{eqnarray}
 F_a^+(x_A,u)=\displaystyle\sum_{n=0}^{\infty}
   f^{(i_1\cdots i_{n+1}j_1\cdots j_n)}(x_A)
   u_{(i_1}^+ \cdots u_{i_{n+1}}^+u_{j_1}^- \cdots u_{j_n)}^-.
   \label{eq:harmonicexp}
\end{eqnarray}
Thus, the hypermultiplet includes infinitely many auxiliary fields in
 addition to physical fields.

We also use the convention to raise and lower the $SU(2)$ indices 
by means of 
$\epsilon_{ij}$ and $\epsilon^{ij}$, 
\begin{eqnarray}
\epsilon_{21}=\epsilon^{12}=1,
\qquad 
\epsilon_{12}=\epsilon^{21}=-1. 
\end{eqnarray}
For instance the scalar fields for hypermultiplet $f_a^i, 
i=1,2$ have the property 
: 
\begin{eqnarray}
f_a^i&=&\epsilon^{ij} f_{aj}, 
\qquad 
f_{ai}=\epsilon_{ij} f_{a}^j, 
\\ 
\bar f_{ai}&=&\epsilon_{ij} \bar f_{a}^j,  
\qquad 
\bar f_{a}^i=\epsilon^{ij} \bar f_{aj},  
\end{eqnarray}
where lower index $a$ denotes fundamental representation 
 in flavor symmetry group.
Therefore the scalar fields for hypermultiplet has the following 
reality property in conformity with our convention of complex conjugate 
in HSF (\ref{eq:realityHSF}) : 
\begin{eqnarray}
\left(f_a^i\right)^*= \overline{ f_{a}^i} \equiv 
\bar f_{ai}. 
\end{eqnarray}
Namely we have $\bar f_{a1}=-\bar f_a^2$, 
$\bar f_{a2}=\bar f_a^1$. 
We shall use $f_a^i$ and its complex 
conjugate field $\bar f_{ai}$ as much as possible instead of 
$\bar f_a^i = \epsilon^{ij} \bar f_{aj}$.

\section{
The Eguchi-Hanson Nonlinear Sigma Model  in HSF 
}\label{sc:EH-HSF}

The massive HK sigma model on Eguchi-Hanson 
manifold \cite{EH} ($T^*{\bf C}P^1$) 
is described in terms of harmonic superfields \cite{Ivanov2} 
integrated over the analytic subspace $d\zeta_A^{(-4)}du$  
\begin{equation}
\!\! S
\!=\!
-\!\!\displaystyle\int\! d\zeta_A^{(-4)}du \left({\widetilde{q_1^+}} D^{++}q_1^+
   +{\widetilde{q_2^+}}D^{++}q_2^+ +V^{++}({\widetilde{q_1^+}} q_2^+ 
   -{\widetilde{q_2^+}}q_1^+ +\xi^{++})\right) \label{eq:EHhss}
\end{equation}
where the covariant derivative $D^{++}$ defined in 
(\ref{covdhss}) contains the central charge $Z$ 
 satisfying the following eigenvalue 
 equation 
\begin{eqnarray}
 Z q_a^+(\zeta_A,u) = \frac{\mu}{2} q_a^+(\zeta_A,u) .
\end{eqnarray}
This mass parameter can be attributed to the 
Sherk-Schwarz reduction from six 
dimensions\cite{DRHSF}
 : 
$Z = -i(\partial_5+i \partial_6)$. 

The Lagrangian (\ref{eq:EHhss}) is invariant under $O(2)$ gauge 
 transformation 
\begin{eqnarray}
 \delta q_1^+(\zeta_A,u) &=& -\lambda(\zeta_A,u) q_2^+(\zeta_A,u), 
\\
\qquad 
 \delta q_2^+(\zeta_A,u) 
&
=
&
 \lambda(\zeta_A,u) q_1^+(\zeta_A,u), 
\\
  \delta V^{++}(\zeta_A,u) 
  &=& D^{++} \lambda(\zeta_A,u). \label{gthss3}
\end{eqnarray}

Similarly to the Grassmann expansion of hypermultiplets 
(\ref{eq:exp1}), the vector multiplet $V(\zeta_A, u)$ can also be 
expanded into infinitely many components when expanded 
in powers of Grassmann numbers $\theta$. 
These components can then be expanded into power series in harmonic 
variables $u_i^{\pm}$. 
However, we can exploit the gauge transformation (\ref{gthss3}) 
to eliminate most of the auxiliary components in powers of Grassmann 
number $\theta$ and also in powers of harmonic variables $u_i^{\pm}$ 
in the vector multiplet. 
After eliminating infinitely many auxiliary fields by the gauge 
transformations, we obtain a gauge fixing 
\begin{eqnarray}
 V_{\rm WZ}^{++}(x_A, \theta^{\pm},u) 
 &=&  \theta^+ \theta^+ {\bar{M}}_v(x_A) 
 +{\bar{\theta}}^+ {\bar{\theta}}^+ M_v(x_A)
 -2 i \theta^+ \sigma^\mu  {\bar{\theta}}^+ V_\mu(x_A)
                       \nonumber \\
& + &\sqrt{2} \theta^+\theta^+{\bar{\theta}}^+ {\bar{\lambda}}^i(x_A) u_i^-
                      + \sqrt{2} {\bar{\theta}}^+ {\bar{\theta}}^+
                        \theta^+ \lambda^i(x_A) u_i^{-}
                       \nonumber \\
                   & + &\theta^+ \theta^+ {\bar{\theta}}^+ {\bar{\theta}}^+
                      D_v^{(ij)}(x_A)u_{(i}^- u_{j)}^-, 
                      \label{eq:exp2}
\end{eqnarray}
which is called the Wess-Zumino gauge and is denoted by the suffix 
WZ. 
As a result, the remaining fields 
$M_v(x_A),~V_\mu(x_A),~\lambda^i(x_A)$ in (\ref{eq:exp2}) 
are physical fields except $D_v(x_A)^{(ij)}$, 
if there is a kinetic term for vector multiplet. 
The field $D_v(x_A)^{(ij)}$ is the usual SUSY auxiliary field. 
However, we will use here a vector multiplet with no kinetic term. 
Therefore we will eventually eliminate all these component fields in the 
vector multiplet giving rise to constraints for hypermultiplets. 

After integrating Grassmann variables 
 and the harmonic variables, 
and eliminating infinitely many auxiliary fields of the hypermultiplet 
expanded in powers of harmonic variables $u_i^{\pm}$, and taking 
the Wess-Zumino gauge for the vector multiplet in HSF, 
we obtain the bosonic part of the action as 
\begin{eqnarray}
&&{\cal L}_{\rm boson}^{\rm HSF}
  \nonumber \\
& 
= 
&
 - \left(\partial_A^\mu f_1^i + V^\mu f_2^i\right)
  \left(\partial_\mu^A {\bar{f}}_{1i} + V_\mu {\bar{f}}_{2i}\right) 
 -  \left(\partial_A^\mu f_2^i - V^\mu f_1^i\right)
  \left(\partial_\mu^A {\bar{f}}_{2i} - V_\mu {\bar{f}}_{1i}\right) 
  \nonumber \\ & & 
-\frac{1}{2}\left({\bar{M}}_v {\bar{f}}_{1i}
-\frac{\bar\mu}{2} {\bar{f}}_{2i}\right)
 \left(M_v f_{1}^{i} - \frac{\mu}{2} f_{2}^{i}\right)
-\frac{1}{2}\left({\bar{M}}_v {\bar{f}}_{2i} 
+ \frac{\bar\mu}{2} {\bar{f}}_{1i}\right)
 \left(M_v f_{2}^{i} + \frac{\mu}{2} f_{1}^{i}\right)
  \nonumber \\ & & 
-\frac{1}{2}\left(M_v {\bar{f}}_{1i} 
+ \frac{\mu}{2} {\bar{f}}_{2i}\right)
  \left({\bar{M}}_v f_{1}^{i} 
+ \frac{\bar\mu}{2} f_{2}^{i}\right)
- \frac{1}{2}\left(M_v {\bar{f}}_{2i} 
- \frac{\mu}{2} {\bar{f}}_{1i}\right)
  \left({\bar{M}}_v f_{2}^{i} 
- \frac{\bar\mu}{2} f_{1}^{i}\right) \nonumber \\
 & & 
- \frac{1}{3}D_{v(ij)}(-{\bar{f}}_1^{(i}f_2^{j)}
     +{\bar{f}}_2^{(i}f_1^{j)}+\xi^{(ij)}) 
  \nonumber \\
&
= 
&
{\cal L}_{\rm kin}^{\rm HSF}
+{\cal L}_{\rm constr}^{\rm HSF}
+{\cal L}_{\rm pot}^{\rm HSF},
                      \label{eq:HSFO2action} 
\end{eqnarray}
\begin{eqnarray}
{\cal L}_{\rm kin}^{\rm HSF}
&
= 
&
-\partial_\mu^A f_a^i \partial^\mu_A {\bar{f}}_{ai}
+\partial_\mu^A f_a^i \epsilon_{ab} V^\mu {\bar{f}}_{bi}
-\epsilon_{ab}V^\mu f^i_a \partial_\mu^A {\bar{f}}_{bi}
\nonumber \\
&&-V^\mu V_\mu f_a^i{\bar{f}}_{ai},
                      \label{eq:HSFO2kin} 
\end{eqnarray}
\begin{eqnarray}
{\cal L}_{\rm constr}^{\rm HSF}
 &
=
 & 
- \frac{1}{3}D_{v(ij)}(-{\bar{f}}_1^{(i}f_2^{j)}
     +{\bar{f}}_2^{(i}f_1^{j)}+\xi^{(ij)}), 
                      \label{eq:HSFO2constr-lag} 
\end{eqnarray}
where $a=1,2$ denotes fundamental representation
 in $O(2)$ gauge group.
The scalar potential 
$V(f,{\bar{f}})$ is given by 
\begin{eqnarray}
-
{\cal L}_{\rm pot}^{\rm HSF}
&
=
&
V(f,{\bar{f}})
\nonumber \\
&=&
\left(\left|{\mu \over 2}\right|^2+|M_v|^2\right)
\left(f_1^i{\bar{f}}_{1i}+f_2^i{\bar{f}}_{2i}\right)
\nonumber \\
&
+
&
\left({\mu \over 2}{\bar{M}}_v-{\bar{\mu \over 2}}M_v\right)
\left(f_1^i{\bar{f}}_{2i}-f_2^i{\bar{f}}_{1i}\right)
.
                      \label{eq:HSFO2pot} 
\end{eqnarray}
Let us stress once again that 
we adopt a convention for complex conjugation of complex scalar fields  
$
\left(f_a^i\right)^*\equiv \bar f_{ai}=\epsilon_{ij} \bar f_a^j, 
$ 
and use $f_a^i$ and $\bar{f}_{ai}$ to denote a complex conjugate pair. 

There are still auxiliary fields $M_v$ and $V^\mu$ 
and $D_{v (i j)}$ of the 
 vector multiplet.
By changing variables, we can also introduce 
the most frequently used parameterization given by 
 Curtright and Freedman \cite{CF} : four complex fields 
 $\phi^\alpha_i, \alpha=1,2, i=1,2$
\begin{eqnarray}
 \phi_1^\alpha=\frac{1}{\sqrt{2}}(f_1^{2,\alpha} + i f_2^{2,\alpha}),
 ~~~~~
 \phi_2^\alpha=\frac{1}{\sqrt{2}}(f_1^{1,\alpha} + i f_2^{1,\alpha}),
 \label{eq:repar1}
\end{eqnarray}
where $f_a^{i,1}=f_a^i$ and $f_a^{i,2}={\bar{f}}_a^i$. 

The Lagrangian in the HSF 
(\ref{eq:HSFO2action})-(\ref{eq:HSFO2pot}) 
can be related to the component Lagrangian 
(\ref{eq:N1O2boson})-(\ref{eq:N1O2pot}) 
in terms of 
${\cal N}=1$ superfields in the $O(2)$ basis 
in the Wess-Zumino gauge 
by the following identification : 
\begin{equation}
M_v = i \Sigma, 
\qquad
V_\mu={v_\mu \over 2}, 
\end{equation} 
and fields $f^i_a$ of HSF 
can be identified with ${\cal N}=1$ fields $\tilde \phi^a, \tilde \chi_a$ 
in the $O(2)$ basis (\ref{eq:N1O2boson})-(\ref{eq:N1O2pot}) 
as 
\begin{equation}
f^1_a
=
\left(\tilde \phi^a\right)^*, 
\qquad
f^2_a=\tilde \chi_a. 
\label{eq:HSF-N1O2}
\end{equation} 
The Fayet-Iliopoulos parameters $\xi^{(ij)}$ 
in HSF are identified with Fayet-Iliopoulos parameters 
$c, b, b^*$ 
in the ${\cal N}=1$ 
superfield formalism as 
\begin{equation}
\xi^{11}=-i b^*, 
\qquad
\xi^{22}
=
i b, 
\qquad
\xi^{12}=
\xi^{21}
=
{ic\over 2} .
\end{equation} 
The  auxiliary fields $D_{v(ij)}$ 
in HSF are identified with 
 the auxiliary fields $D, F_{\Sigma}, F_{\Sigma}^*$ 
in the ${\cal N}=1$ 
superfield formalism as 
\begin{equation}
D_{v(11)}=3i F_{\Sigma}^*, 
\qquad
D_{v(22)}=-3i F_{\Sigma}, 
\qquad
D_{v(12)}=-{3i D \over 2}
. 
\end{equation} 
These results are in conformity with the reality property of 
the Fayet-Iliopoulos parameters $\xi^{(ij)}$ and the 
  auxiliary fields $D_{v(ij)}$ 
in HSF and those in ${\cal N}=1$ superfield formalism  
\begin{equation}
\xi^{(ij)}=\epsilon^{ik}\epsilon^{jl}\left(\xi^{(kl)}\right)^* 
, 
\qquad
D_{v(ij)}=\epsilon_{ik}\epsilon_{jl}\left(D_{v(kl)}\right)^* 
. 
\end{equation} 
\begin{equation}
b \in {\bf C}, \quad c \in {\bf R}
, 
\qquad
F_{\Sigma} \in {\bf C}, \quad D \in {\bf R}
. 
\end{equation} 
The identification (\ref{eq:HSF-N1O2}) implies 
that the complex scalar fields $f_a^1$ belong to 
anti-chiral scalar superfields, and  $f_a^2$ to 
chiral scalar superfields. 
The suffix $a$ denotes fundamental representation of the 
gauge group $O(2)$. 

The complex fields $\phi_i^\alpha$ in the Curtright-Freedman 
basis (\ref{eq:repar1}) are more directly related to 
the complex scalar fields of the ${\cal N}=1$ superfields 
$\phi_i, \chi_i$ in the $U(1)$ basis in 
(\ref{eq:N1U1boson})-(\ref{eq:N1U1pot}) as 
\begin{eqnarray}
\!\!
\phi_1{}^1 
&
=
\!\!
& 
{1 \over \sqrt2}\left( f_1^{2}+ i  f_2^{2}\right)
=
\!\!
{1 \over \sqrt2}\left( \tilde \chi_{1}+i\tilde \chi_{2}\right)
=
\chi_1, 
\\
\!\!
\phi_2{}^1
&
=
\!\!
& 
{1 \over \sqrt2}\left( f_1^{1}+ i  f_2^{1}\right)
=
\!\!
{1 \over \sqrt2}\left( \tilde \phi_{1*}+i\tilde \phi_{2*}\right)
=\left(\phi_1\right)^*, 
\\
\!\!
\phi_1{}^2
&
=
\!\!
& 
{1 \over \sqrt2}\left(\bar f_1^{2}+ i \bar f_2^{2}\right)
=
\!\!
{1 \over \sqrt2}\left(- \bar f_{11}-i\bar f_{21}\right)
=
\!\!
{1 \over \sqrt2}\left(- \tilde \phi^{1}-i\tilde \phi^{2}\right)
=\chi_2, 
\\
\!\!
\phi_2{}^2
&
=
\!\!
& 
{1 \over \sqrt2}\left(\bar f_1^{1}+ i \bar f_2^{1}\right)
=
\!\!
{1 \over \sqrt2}\left( \bar f_{12}+i\bar f_{22}\right)
=
\!\!
{1 \over \sqrt2}\left( \tilde \chi_1^{*}+i\tilde \chi_2^{*}\right)
=\left(\phi_2\right)^* .
\label{eq:CF-N1ident}
\end{eqnarray} 
Therefore the complex scalar fields $\phi_1^i$ are identified as those of 
chiral scalar superfield, and  $\phi_2^i$ are identified as those of 
anti-chiral scalar superfield. 
We also notice that all the complex fields 
$\phi_i^\alpha, i=1,2, \alpha=1,2$ 
in the Curtright-Freedman basis have 
$U(1)$ charge $-1$ in conformity with the charge 
assignment obtained in the model constructed by 
the tensor calculus for supergravity.\cite{AFNS} 
This supergravity model shows that our model can be embedded 
into supergravity. 
Moreover it explicitly demonstrates that our model can be 
extended to a model in five dimensions. 

The equation of motion for the gauge field $V_\mu$ gives 
\begin{eqnarray}
V_\mu
=
{  \epsilon_{ab} 
\left(\partial_\mu^A f_a^i {\bar{f}}_{bi}
- f^i_a \partial_\mu^A {\bar{f}}_{bi}\right)
\over 
2 f_a^i{\bar{f}}_{ai}
}.
\end{eqnarray} 
After eliminating the vector field $V^\mu$, 
we obtain the kinetic term for the scalar fields 
$f_a^i$ in the hypermultiplets 
\begin{eqnarray}
{\cal L}_{\rm kin}^{\rm HSF} &=& 
-\partial_A^\mu f_1^i \partial_\mu^A {\bar{f}}_{1i}
-\partial_A^\mu f_2^i \partial_\mu^A {\bar{f}}_{2i}
+\frac{(f_2^i \overset{\leftrightarrow}{\partial^\mu_A}{\bar{f}}_{1i} 
-f_1^i \overset{\leftrightarrow}{\partial^\mu_A}{\bar{f}}_{2i})^2}
{4 (f_1^i {\bar{f}}_{1i} + f_2^i {\bar{f}}_{2i})}. 
\label{bosonachss}
\end{eqnarray} 
Integrating over 
scalar $M_v$ and the 
auxiliary fields $D_{v (i j)}$ in the vector multiplet, 
we obtain constraints  
\begin{eqnarray}
 -{\bar{f}}_1^{(i}f_2^{j)}+{\bar{f}}_2^{(i}f_1^{j)}+\xi^{(ij)}
 =0 . \label{consthss}
\end{eqnarray}
This constraint makes the target space 
of the massive nonlinear sigma model into the 
 Eguchi-Hanson manifold.\cite{ANNS}
In the case of massless 
(without potential) model,
the target metric for the four independent real 
 bosonic fields has been shown to be just the Eguchi-Hanson 
 metric. \cite{CF,ivanov-EH,valent} 

The equation of motion for scalar field $M_v$ gives 
\begin{eqnarray}
 M_v={\mu \over 2}
{\epsilon_{ab}f_a^i\bar f_{bi} \over f_c^i\bar f_{ci}}
 \label{eq:MEOM}
\end{eqnarray}
where the flavor indices are summed.
Integrating over $M_v$ gives the potential term as 
\begin{equation}
 V(f_1,f_2) = \left|\frac{\mu}{2}\right|^2
               \frac{1}{f_1^i {\bar{f}}_{1i} + f_2^i {\bar{f}}_{2i}}
               \left\{-
|f_1^i {\bar{f}}_{2i} - f_2^i {\bar{f}}_{1i}|^2
           +(f_1^i {\bar{f}}_{1i} + f_2^i {\bar{f}}_{2i})^2 \right\}. 
               \label{pothss}
\end{equation} 
The parameters $\xi^{(ij)}$ have mass dimensions two and represent 
the scale of the curvature of the target manifold.

The bosonic action becomes in the Curtright-Freedman basis as 
\begin{eqnarray}
&\!&\!{\cal L}_{\rm boson} = 
-\partial_A^\mu \phi_1\partial^A_\mu {\bar{\phi}}_1
-\partial_A^\mu \phi_2\partial^A_\mu {\bar{\phi}}_2
-\frac{(\phi_1  \overset{\leftrightarrow}{\partial^\mu_A}
                            {\bar{\phi_1}}
                            +\phi_2  \overset{\leftrightarrow}{\partial^\mu_A} 
                            {\bar{\phi_2}})^2}{4(|\phi_1|^2+|\phi_2|^2)}
                   \nonumber \\
  &\!   -&\!\frac{\mu^2}{4(|\phi_1|^2+|\phi_2|^2)}
              {\bigg(}-(\phi_1 \sigma^3 \bar{\phi}_1
               +\phi_2 \sigma^3 \bar{\phi}_2)^2 
               +(|\phi_1|^2+|\phi_2|^2)^2 {\bigg )}
.  
\label{eq:CFaction}
\end{eqnarray}

\section{
Nonlinear Sigma Model in Independent Fields : 
Spherical Coordinates and Gibbons-Hawking Parameterization 
}

Here we shall describe the model in terms of independent 
fields in several parameterizations by solving the constraints 
(\ref{consthss}). 
In the following we shall take 
\begin{equation}
\xi^{(12)} \equiv -i\xi, \qquad \xi^{(11)}=\xi^{(22)}=0  
\end{equation}
for simplicity. 
Then the constraints (\ref{consthss}) become 
\begin{equation}
  |\phi_1|^2-|\phi_2|^2 = 2\xi,  \qquad 
  \phi_1^*\phi_2 = \phi_2^*\phi_1 = 0.
\label{CFconst}
\end{equation}

It is convenient to introduce independent fields $z^\alpha, 
\bar z^\alpha, \; \alpha =1, 2$ through 
the following Ansatz \cite{AF2,valent} 
\begin{eqnarray}
 \phi_1^\alpha = g(r){z^\alpha \over \sqrt{r}},
\qquad 
 \phi_2^\alpha = 
 f(r)i\sigma_2^{\alpha\beta}{{\bar{z}}^\beta \over \sqrt{r}}, 
 \label{eq:repar2}
\end{eqnarray} 
\begin{eqnarray}
r=z^1{\bar{z}}^1+z^2{\bar{z}}^2
,
\end{eqnarray} 
where $z^\alpha$ are complex fields satisfying 
\begin{eqnarray}
 \phi_1^1\phi_2^2-\phi_1^2\phi_2^1= - z^1{\bar{z}}^1-z^2{\bar{z}}^2. 
 \label{eq:phi_z}
\end{eqnarray} 
The real functions $f(r)$ and $g(r)$ 
are uniquely determined 
by the constraints (\ref{consthss}) and (\ref{eq:phi_z}) as 
\begin{eqnarray}
f(r)^2 = 
-\xi + \sqrt{r^2 + \xi^2}
,
\qquad  
g(r)^2 = 
\xi + \sqrt{r^2 + \xi^2}
.
\end{eqnarray}
The action can be described without constraint by the independent 
 complex fields $z^\alpha$. These fields $z^\alpha$ are invariant under the 
$O(2) (U(1))$ gauge transformations in (\ref{gthss3})  
which is used to take the quotient of  
the target manifold. 

Another useful parameterization of the model is given by the 
spherical coordinates which are invariant under the $U(1)$ 
gauge transformations 
\begin{eqnarray}
 & z^1 = \sqrt{r}\cos \frac{\Theta}{2} \exp \frac{i}{2} (\Psi + \Phi),&
         \label{eq:repar3} \\
 & z^2 = \sqrt{r}\sin \frac{\Theta}{2} \exp \frac{i}{2} (\Psi - \Phi),&
         \label{eq:repar4} \\
 & 0 \le r \le \infty,
\quad
0 \le \Theta \le \pi,
\quad
0 \le \Phi \le 2\pi,
\quad
0 \le \Psi \le 2\pi. & 
\end{eqnarray}
\begin{eqnarray}
\!\!\!\!\!\!  f_1^1&=&
  {\phi_2^1-\bar \phi_1^2 \over \sqrt2}
=  -{g(r)-f(r) \over \sqrt{2r}}\bar z^2
=  -{g(r)-f(r) \over \sqrt2}
  \sin(\frac{\Theta}{2}){\rm e}^{\frac{i}{2}(-\Psi+\Phi)}, 
  \\
\!\!\!\!\!\! f_2^1&=&
  {\phi_2^1+\bar \phi_1^2 \over i\sqrt2}
=  {g(r)+f(r) \over i\sqrt{2r}}\bar z^2
=  {g(r)+f(r) \over i\sqrt2}
\sin(\frac{\Theta}{2}){\rm e}^{\frac{i}{2}(-\Psi+\Phi)}
, 
  \\
\!\!\!\!\!\! f_1^2&=&
  {\phi_1^1+\bar \phi_2^2 \over \sqrt2}
=  {g(r)-f(r) \over \sqrt{2r}} z^1
={g(r)-f(r) \over \sqrt2}
\cos(\frac{\Theta}{2}){\rm e}^{\frac{i}{2}(\Psi+\Phi)}
, 
  \\
\!\!\!\!\!\! f_2^2&=&  {\phi_1^1-\bar \phi_2^2 \over i\sqrt2}
=  {g(r)+f(r) \over i\sqrt{2r}} z^1
={g(r)+f(r) \over i\sqrt2}
\cos(\frac{\Theta}{2}){\rm e}^{\frac{i}{2}(\Psi+\Phi)}
.
\label{Constsolve}
\end{eqnarray}
\begin{eqnarray}
  \phi_1^1
&
=
&
g(r)\cos(\frac{\Theta}{2})\exp(\frac{i}{2}(\Psi+\Phi)), 
\\
\phi_1^2
&
=
&
g(r)\sin(\frac{\Theta}{2})\exp(\frac{i}{2}(\Psi-\Phi)), 
\\
\phi_2^1
&
=
&
f(r)\sin(\frac{\Theta}{2})\exp(-\frac{i}{2}(\Psi-\Phi)), 
\\
\phi_2^2
&
=
&
-f(r)\cos(\frac{\Theta}{2})\exp(-\frac{i}{2}(\Psi+\Phi)) .
\end{eqnarray}
The bosonic action becomes in the spherical coordinates as 
\begin{eqnarray}
{\cal L}_{\rm boson}
  &= & 
{1 \over 2\sqrt{\xi^2 + r^2} }
\Biggl[
-\partial^A_\mu r \partial_A^\mu r 
- \left(r^2+ \xi^2\right) \partial^A_\mu \Theta \partial_A^\mu \Theta 
 \nonumber \\
   &{}&  
- \left(r^2+ \xi^2 \sin^2 \Theta\right) \partial^A_\mu \Phi 
\partial_A^\mu \Phi 
- r^2 \partial^A_\mu \Psi \partial_A^\mu \Psi 
- 2 r^2 \cos \Theta \partial^A_\mu \Phi \partial_A^\mu \Psi 
 \nonumber \\
   &{}&  
- |\mu|^2 \left(r^2+\xi^2 \sin^2 \Theta\right) \Biggr] .
\label{eq:sph-coo-action}
\end{eqnarray}

It is also useful to change variables into the following 
 parameterization appropriate to describe 
the Gibbons-Hawking multi-center metric~\cite{GH}
\begin{eqnarray}
 X^1 &=& r \sin \Theta \cos \Psi, \label{repar5} \\
 X^2 &=& r \sin \Theta \sin \Psi, \label{repar6} \\ 
 X^3 &=& \sqrt{r^2 + \xi^2} \cos \Theta, \label{repar7} \\
 \varphi &=& \Phi + \Psi. \label{repar8} 
 \end{eqnarray}
By using the parameterisation 
 (\ref{eq:repar1})-(\ref{eq:repar4}) and (\ref{repar5})-(\ref{repar8}),
 the bosonic part of the action (\ref{bosonachss}) can be rewritten as 
\begin{equation}\label{nlsm_lag}
 {\cal L} 
 = -{1\over 2}\left\{U\partial_\mu{\bf X}\cdot
 \partial^\mu {\bf X} + U^{-1}{\cal D}_\mu \varphi {\cal D}^\mu
 \varphi + \mu^2 U^{-1}\right\},
\end{equation}
where
 ${\cal D}_\mu\varphi 
= \partial_\mu\varphi + {\bf A} \cdot 
 \partial_\mu{\bf X}$
 and 
\begin{eqnarray}
 {\mbox{\boldmath $\nabla$}} \times {\mbox{\bf A}} 
 = \mbox{\boldmath $\nabla$} U. 
\label{eq:monopole} 
\end{eqnarray}
The harmonic function $U$ can be described
\begin{equation}
 U =  
{1\over2}\left[{1\over |{\bf X}- \xi{\bf n}|} + 
 {1\over |{\bf X}+ \xi{\bf n}|}\right]\, ,
 \label{eq:harmonic_func}
\end{equation}
where ${\bf n}$ is a unit three vector, which is given by 
 ${\bf n}=(0,0,1)$. 
${\bf A}$ is a potential whose solution is given as
\begin{eqnarray}
 A_1 
 &\!=&\! {1 \over 2}
  \left\{
    \frac{X^2} 
    {|{\bf X} - \xi {\bf n}| (X^3 - \xi + |{\bf X} - \xi {\bf n}|)}
   +\frac{X^2} 
   {|{\bf X} + \xi {\bf n}| (X^3 + \xi + |{\bf X} - \xi {\bf n}|)}
   \right\}, \nonumber \\
 A_2 
 &\!=&\! {1 \over 2}
 \left\{
    \frac{-X^1} 
    {|{\bf X} - \xi {\bf n}| (X^3 - \xi + |{\bf X} - \xi {\bf n}|)}
   +\frac{-X^1} 
   {|{\bf X} + \xi {\bf n}| (X^3 + \xi + |{\bf X} + \xi {\bf n}|)}
   \right\}, \nonumber \\  
 A_3 &\!=&\! 0.
\end{eqnarray}
It is found that the target metric of the action (\ref{nlsm_lag}) 
is just the Eguchi-Hanson 
 metric.\cite{CF,ivanov-EH,valent}

\section{
BPS Equation and Domain Wall Solution
}\label{sc:HSF-BPSeq}
In this Sec., we give the BPS domain wall solution in our model. 
In the following the complex mass parameter $\mu$ is taken to be real 
for simplicity. 
By requiring that the fermions 
conserve half of SUSY we obtain 
the BPS equations in HSF 
\begin{eqnarray}
 (M_v + V_2) f_2^1 + \left(\frac{\mu}{2} + \partial_2^A\right) f_1^1 &=& 0, 
   \label{eq:BPSeqhss5} \\
 (M_v - V_2) f_2^2 + \left(\frac{\mu}{2} - \partial_2^A\right) f_1^2 &=& 0, 
   \label{eq:BPSeqhss6} \\
 -(M_v + V_2) f_1^1 + \left(\frac{\mu}{2} + \partial_2^A\right) f_2^1 &=& 0, 
   \label{eq:BPSeqhss7} \\
 -(M_v - V_2) f_1^2 + \left(\frac{\mu}{2} - \partial_2^A\right) f_2^2 &=& 0.
   \label{eq:BPSeqhss8}
\end{eqnarray}

BPS wall solution should approach the supersymmetric discrete 
 vacua as $y\rightarrow \pm\infty$. 
 From the trivial solution of BPS equation (translational invariant 
 solution), we find \cite{ANNS} that there are only two vacua 
: $(r,\Theta)=(0,0),(0,\pi)$ in terms of the spherical coordinates 
(\ref{eq:repar3}), (\ref{eq:repar4}). 
Another way of understanding these vacua is to observe from 
Eq.(\ref{eq:sph-coo-action}) that these two 
points are the minima of the scalar potential 
\begin{eqnarray}
V (r, \Theta, \Phi, \Psi) =
{ |\mu|^2 \left(r^2+\xi^2 \sin^2 \Theta\right)   
 \over 2\sqrt{\xi^2 + r^2} }
\end{eqnarray}
with vanishing vacuum energy 
\begin{eqnarray}
V(r=0, \Theta=0)=V(r=0, \Theta=\pi)=0. 
\end{eqnarray}

\begin{figure}[t]
\begin{center}
\leavevmode
  \epsfxsize=6.0cm
  \epsfysize=4.0cm
  \epsfbox{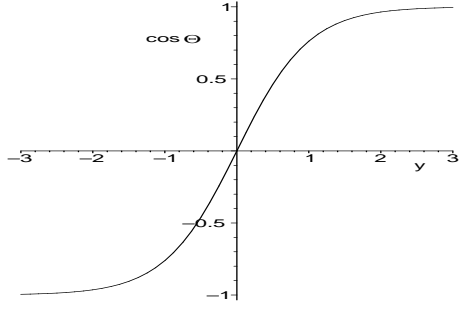} \\
\caption{
BPS domain wall solution of field $\cos \theta$ as a function of $y$ with 
$y_0=0$.}
\label{fig:BPSwall}
\end{center}
\end{figure}

Therefore we consider the domain wall solution connects these vacua,
and we can expect that $\Theta$ has nontrivial configuration. 
After some algebra, we obtain four independent 
differential equations in terms of the spherical coordinates 
\cite{ANNS} 
\begin{eqnarray}
r' &=& \mu \cos \Theta \cdot r, \qquad r\cdot \Psi' = 0, 
\label{fulleq1} \\
\Theta' &=& - \mu \sin \Theta, \qquad \sin\Theta\cdot\Phi' = 0. 
\label{fulleq2}
\end{eqnarray}
The boundary condition of $r=0$ at $y=-\infty$ 
dictates 
the solution 
of (\ref{fulleq1}) to be 
$r=0$ and  $\Psi = 0$. 
The other two equations in (\ref{fulleq2}) 
gives a nontrivial dependence in $y$ 
resulting in the following BPS solutions 
\begin{eqnarray}
 \Theta=\arccos[\tanh\mu (y + y_0)], 
\qquad
\Phi = \varphi_0, 
\label{eq:solhss}
\end{eqnarray}
where $y_0$ and $\varphi_0$ are real constants: 
$y_0$ determines the position of the domain wall 
along $y$ direction and $\varphi_0$ corresponds to the 
Nambu-Goldstone (NG) mode of $U(1)$ isometry of target space. 
The BPS wall solution is illustrated in Fig.\ref{fig:BPSwall}. 

Our BPS wall solution 
is obtained from ${\cal N}=2$ SUSY theory in four dimensions. 
However, subsequent study\cite{AFNS} 
revealed that our model can be extended to an 
 ${\cal N}=2$ SUSY theory in five dimensions. 
 In fact we obtain precisely the same BPS solution as in 
(\ref{eq:solhss}) by taking the limit of vanishing gravitational 
coupling in the BPS wall solution in five-dimensional 
supergravity.\cite{AFNS} 
Therefore we can now use our BPS wall solution as a starting point for 
an interesting phenomenology for a unified model : 
our four-dimensional spacetime being 
a wall in higher dimensional spacetime following the 
brane-world scenario. 

In terms of harmonic superfields (\ref{eq:EHhss}) 
and their bosonic components (\ref{bosonachss}), 
the BPS solution is given by 
\begin{eqnarray}
 q_1^+=f^i_1 u_i^+ &=&\sqrt{\frac{\xi}{2}}{\rm e}^{\frac{i}{2}\varphi_0}
                       \left(\begin{array}{c}
                           -\sqrt{1-\tanh(\mu (y+y_0))} u_1^+ \\     
                            \sqrt{1+\tanh(\mu (y+y_0))} u_2^+  
                       \end{array}
                      \right), \\
 q_2^+=f^i_2 u_i^+ &=&-i\sqrt{\frac{\xi}{2}}{\rm e}^{\frac{i}{2}\varphi_0}
                      \left(\begin{array}{c}
                            \sqrt{1-\tanh(\mu (y+y_0))} u_1^+ \\     
                            \sqrt{1+\tanh(\mu (y+y_0))} u_2^+  
			    \end{array}
                      \right). 
\end{eqnarray}

In terms of the fields in the Curtright-Freedman basis 
(\ref{eq:CFaction}), 
the BPS wall solution is given by 
\begin{eqnarray}
\phi_1^1&=&\sqrt{\xi(1+\tanh\mu(y+y_0))}{\rm e}^{\frac{i}{2}\varphi_0},
     \\
\phi_1^2&=&\sqrt{\xi(1-\tanh\mu(y+y_0))}{\rm e}^{-\frac{i}{2}\varphi_0},
     \\
\phi_2^1&=&\phi_2^2=0.
\end{eqnarray}

The BPS wall solution in the Gibbons-Hawking multi-center metric 
parameterization (\ref{nlsm_lag}) is given by 
\begin{eqnarray}
 X^1 &=& X^2 = 0, \\
X^3 &=& \xi \tanh\mu(y + y_0), 
\\ 
\varphi& = & \varphi_0.
\end{eqnarray}

\section*{Acknowledgements}
\addcontentsline{toc}{section}{Acknowledgements}

One of the authors (NS) thanks Yoshiaki Tanii for useful 
discussion on hypermultiplets in six dimensions. 
This work is supported in part by Grant-in-Aid 
 for Scientific Research from the Japan Ministry 
 of Education, Science and Culture  13640269. 
The work of M.~Naganuma is supported by JSPS Fellowship. 
The work of M.~Nitta is supported by the U.~S. Department
 of Energy under grant DE-FG02-91ER40681 (Task B).

\end{document}